\documentclass[aps,twocolumn]{revtex4}
\usepackage{psfig}
\usepackage{subfigure}
\usepackage{graphicx}

\begin{document}



\title{Pinning-Controllability of Complex Networks}
\author{Francesco Sorrentino${}^{*\ddagger}$, Mario di Bernardo${}^*$, Franco Garofalo${}^*$, Guanrong Chen${}^\dagger$}
\affiliation{${}^*$University of Naples Federico II, Naples 80125,
Italy\\ ${}^\dagger$ City University of Hong Kong, P.R. China
\\ ${}^\ddagger$ Corresponding author. E-mail: {\tt
fsorrent@unina.it}}
\begin{abstract}
We study the problem of controlling a general complex network
towards an assigned synchronous evolution, by means of a pinning
control strategy. We define the pinning-controllability of the
network in terms of the spectral properties of an extended network
topology. The roles of the control and coupling gains as well as of
the number of pinned nodes are also discussed.
\end{abstract}
\maketitle

\section{Introduction}

The need of regulating the behavior of large ensembles of
interacting units is a common feature of many physical, social and
biological networks.  For instance, many regulatory mechanisms have
been uncovered in the context of biological, physiological and
cellular processes, which are fundamental to guarantee the correct
functioning of the whole network \cite{VishBOOK}. Examples include,
in physiology, the synchronous beat of heart cells whose rhythm is
generated by pacemaker cells situated at the sinoatrial node
\cite{PesBOOK}, the control of the respiratory rhythm played by
synaptically coupled pacemaker neurons in the medulla \cite{Ko:Sm99}
and the generation of rhythmic pacemaker currents by networks of
interstitial cells of Cajal in the gastrointestinal tract of mammals
\cite{Th:Ro98}. In social networks, opinion dynamics are often
driven by key-individuals termed as \emph{opinion leaders}
\cite{We:De:Am04}. Other relevant examples, in biology, are the
mechanisms through which cell cycles are controlled \cite{LH} and
synchronized \cite{To:Oi:Cr}.

Understanding the fundamental nature of such regulatory mechanisms
is therefore of utmost importance in Physics and Applied Science.
\emph{Pinning control} has been proposed in the literature
\cite{Pinn1,PinnA} as a fitting model to provide an insight into the
regulatory mechanisms to control lattices and networks of coupled
dynamical systems. The general idea behind pinning control is a
self-feedback action (over a given reference evolution), acting on a
limited subset of the dynamical systems placed at the network nodes.
These nodes, also termed as reference sites or pinned sites, play
the role of network \emph{leaders}/\emph{pacemakers}. Specifically,
a direct control action is active only on these nodes and is
propagated to the rest of the network through the coupling among the
vertices. 

In this paper, we define the new concept of {\em
pinning-controllability} (different from synchronizability) and
derive a new quantity to assess this property for physical networks
of interest. These findings have immediate theoretical and
experimental relevance. By evaluating the pinning-controllability of
a given network, one could decide the effectiveness of a pinning
control strategy in terms of the strength of the required control
action, the number of nodes to be pinned and the effects that a
given topology of feedback connections can have on the entire
network.

The rest of the paper is outlined as follows. In Sec. II we
introduce the pinning control formalism, apt to describing 
dynamical complex networks subject to a decentralized control action. 
In Sec. III a new methodology (based on the Master Stability
Function approach \cite{Pe:Ca}) is introduced to define the
pinning-controllability of a given complex network of interest.
Finally, a validation is presented in Sec. IV, where pinning control
of scale-free complex networks of chaotic oscillators is studied
through numerical simulations.

\section{Pinning control: an overview}

In recent years, synchronization of complex networks of coupled
oscillators has been the subject of intensive
research activity within 
the scientific community. 
A common assumption in such literature is that all the dynamical systems at the network nodes are 
identical, while the problem of synchronization of networks of
non-identical oscillators has received much less attention. In
particular, the Master Stability Function (MSF) approach, introduced
in \cite{Pe:Ca}, has been successfully applied to synchronization of
complex networks in a wide variety of situations (see e.g.
\cite{Pe:Ca,Ni:Mo,Bocc2,report}, to name a few contributions). Note
that the MSF approach is valid under the hypothesis that all the
vector fields at the network dynamical nodes are identical.

In this paper, we study networks in which two different layers of
dynamical nodes coexist: the uncontrolled sites and the reference
(controlled) ones. In particular, the latter plays the role of
leading the whole network towards a given (desired) reference
evolution. 

 Generally, we assume the controlled complex network to be
described by the following set of equations:
\begin{equation}
\frac{dx_i}{dt}=f(x_i)+\sigma \sum_{j=1}^{N} {\mathcal A}_{ij}
(h(x_j)-h(x_i))+ \sigma \sum_{k=1}^{n} \delta(i-c_k) u_i,
\label{eq:net2}
\end{equation}
$i=1,...,N$, representing the behavior of $N$ identical dynamical
systems coupled through the network edges.

The first term on the right side of (\ref{eq:net2}) describes the
state dynamics of the oscillator at each node, $\{ x_i(t),
i=1,...,N\}$, via the nonlinear vector field $f(x_i)$; the second
term represents the coupling among pairs of connected oscillators,
through a generic output function $h(x_i)$, where the coupling gain
$\sigma$ represents the overall strength of the interaction.
Information about the weighed network topology is contained in the
coupling matrix $\mathcal A$, whose entries $\mathcal{A}_{ij}$, are
zero if node $i$ is not connected to node $j \neq i$,  but are
positive if there is a direct influence from node $i$ to node $j$,
with $\mid \mathcal{A}_{ij} \mid$ giving a measure of the strength
of the interaction. In what follows, we assume the matrix $\mathcal
A$ to be irreducible (i.e. the associated
digraph is globally connected)
. The last term on the right-hand side of (\ref{eq:net2}) represents
the pinning control action. This term is present only for $n= p N$
(usually $p \ll 1$) \emph{pinned} nodes in the network, identified
by the set $C=\{c_1,c_2,...,c_n\}$. As commonly assumed in {\em
pinning control} schemes, such nodes play the role of leading the
others towards some desired reference evolution, say $s(t)$. Note
that the control input $u_i= \varphi(x_i,s,t)$ has a direct
influence only on the nodes belonging to the set of the reference
sites $C$.

In what follows, we consider two different strategies for choosing
the pinned nodes: (i) \emph{Random pinning}: The $n$ pinned nodes
are randomly selected with uniform probability from the set of all
the network vertices. (ii) \emph{Selective pinning}: The $n$ pinned
nodes are firstly sorted according to a certain property at the
network vertices (for instance, the vertex degree or betweenness
centrality), then the nodes to be pinned are chosen in that
particular order.

Hereafter, following \cite{PinnA}, we choose the control input $u_i$
to be generated by a simple state-feedback law with respect to the
reference evolution $s(t)$, which is assumed to satisfy
$\frac{ds}{dt}=f(s)$. Thus, we set $u_i=\kappa_i(s-x_i)$ at every
pinned node $c_i$, $i=1,...,n$, where $\kappa_i$ is the control
gain acting on node $c_i$. 

Eq. (\ref{eq:net2}) can be rewritten as
\begin{equation}
\frac{dx_i}{dt}=f(x_i)-\sigma \sum_{j=1}^{N} {\mathcal L}_{ij}
h(x_j)- \sigma \sum_{k=1}^{n} \delta(i-c_k) \kappa_i(x_i-s),
 \label{eq:net3}
\end{equation}
$i=1,...,N$, where the elements of the \textit{Laplacian} matrix
$\mathcal{L}$, are as follows: $\mathcal{L}_{ij}=-\mathcal{A}_{ij}$
if $i \neq j$ and $\mathcal L_{ii}= \sum_{j \neq i}
\mathcal{A}_{ij}$ $\forall i$. Denote by $\{\lambda_i=
\lambda_{i}^{r} + j \lambda_{i}^{i}\}$ the set of eigenvalues of
$\mathcal{L}$ and assume they are ordered in such a way that
$\lambda_{1}^r \leq \lambda_{2}^r \leq \cdots \leq \lambda_{N}^r$.

Note that in the specific case where $\mathcal{C}= \emptyset$, i.e.
no reference sites are present in the network, it is possible to
define the \emph{synchronization manifold},
$\mathcal{S}:=\{x_1=x_2=...=x_N\}$, which is an invariant set for
system (\ref{eq:net3}) \cite{Pe:Ca}. In this case, according to the
choice of the coupling gain $\sigma$, all the oscillators in the
network may settle over a solution belonging to $\mathcal{S}$ (not
known {\em a priori}) and the network becomes synchronized.
Interestingly, in \cite{Pe:Ca,Bocc2}, the stability of the manifold
$\mathcal{S}$ was related to $R^N=\frac{\lambda_{N}^r}{\lambda_2^r}$
and $M^N= \max_j \lambda_j^i $, suggesting an evaluation of the
\textit{synchronizability} of a generic network topology in terms of
its spectrum. In particular, a network is said to be more (or less)
synchronizable according to the width of the range of values of the
coupling gain $\sigma$ for which synchronization is attained.
%
%

A different problem consists, instead, in driving a network of
coupled dynamical systems towards a desired evolution determined
{\em a priori}, by means of an appropriate control action. This is
the case where $\mathcal{C} \neq \emptyset$, indicating the presence
of some closed-loop equations within the set $\mathcal{C}$ in (\ref{eq:net3}). 
%
Note that when $\mathcal{C} \neq \emptyset$, the evolution
$x_1(t)=x_2(t)=...=x_N(t)=s(t)$ becomes the only admissible solution
in $\mathcal{S}$ for the controlled network described by
(\ref{eq:net3}). Thus, it is important to study the stability of
such a solution, which corresponds in this context to the model
reference chosen for the network.

It is worth noting here that the Master Stability Function approach
cannot be applied directly to network (\ref{eq:net3}), because of
the
presence 
of inhomogeneous dynamics at the controlled/uncontrolled
nodes. 
In the next section we will provide an extension of the MSF approach
to the case of dynamical networks subject to pinning control, 
described by the set of equations (\ref{eq:net3}).

\section{Evaluating the pinning-controllability of complex networks}

In this section, we propose for the first time that the approach
presented in \cite{Pe:Ca} can be extended to define and assess the
{\em pinning-controllability} of a given network of interest. In
particular, given a network described by (\ref{eq:net3}), we define
the network pinning-controllability in terms of the values of the
coupling gain $\sigma$ and the control gains $\kappa_i$,
$i=1,...,n$, in (\ref{eq:net3}), needed in order for the network to
achieve the desired evolution $\frac{ds}{dt}=f(s)$. We shall seek to
assess the pinning-controllability of network (\ref{eq:net3}) by
considering an extended network of $N+1$ dynamical systems $y_i$,
where $y_i=x_i$ for $i=1,2,..,N$ and $y_{N+1}=s$. In so doing, we
assume that the desired common evolution $s$ is given by the state
evolution of an extra virtual vertex, which is added to the original
network. Then, we can rewrite (\ref{eq:net3}) as
\begin{equation}
\frac{dy_i}{dt}=f(y_i)-\sigma \sum_{j=1}^{N+1} {\mathcal M}_{ij}
h(y_j), \quad i=1,2,...,N+1, \label{eq:net}
\end{equation}
where $\mathcal{M}= \{\mathcal{M}_ {ij} \}$ is an
$(N+1)$-dimensional square matrix, with $\mathcal{M}_{ij}=$
\begin{eqnarray}
{\small\small\small{ \pmatrix{ \mathcal{L}_{11}+ \xi_1 \kappa_1 &
\mathcal{L}_{12} & \cdots & \mathcal{L}_{1N}  & -\xi_1 \kappa_1 \cr
\mathcal{L}_{21} & \mathcal{L}_{22}+\xi_2 \kappa_2 & \cdots &
\mathcal{L}_{2N}  & -\xi_2 \kappa_2\cr \vdots & &\ddots & & \vdots
\cr \mathcal{L}_{N1} & \mathcal{L}_{N2} & \cdots & \mathcal{L}_{NN}+
\xi_N \kappa_N & -\xi_N \kappa_N\cr 0 & 0 & \cdots & 0 & 0 }}}
\nonumber
\end{eqnarray}
and $\xi_i=\sum_{k=1}^{n} \delta(i-c_k)$.

Note that $\mathcal{M}$ is an asymmetric zero row-sum matrix, with
positive values along the main diagonal. At first, let us assume the
matrix $\mathcal A$ to be symmetric, which ensures the matrix
$\mathcal{M}$ to be diagonalizable \footnote{In fact the spectrum of
$\mathcal{M}$ can be decomposed in the spectrum of its minor
containing its first $N$ rows and first $N$ columns, which is a
symmetric matrix, plus one zero eigenvalue.}. Then, let  $\{ \mu_i=
\mu_{i}^{r} + j \mu_{i}^{i}\}$ be the eigenvalues of $\mathcal{M}$
and assume they are ordered in such a way that $\mu_{1}^r \leq
\mu_{2}^r \leq \cdots \leq \mu_{N}^r$. By graph theory and linear
algebra, we have $\mu_i^r \geq 0$ $\forall i$ \cite{ChungBOOK}.
Moreover, $\mu_1$ is the only null eigenvalue of $\mathcal{M}$.

Now, the arguments of the Master Stability Function 
approach \cite{Pe:Ca} immediately apply to (\ref{eq:net}). In
particular, by making use of the MSF, we can investigate the
stability of the reference evolution,
$y_1(t)=y_2(t)=...=y_N(t)=y_{N+1}(t)$, which can be evaluated in
terms of the stability of $N+1$ independent blocks in the parameter
$\alpha=\sigma \mu_i$, $i=1,...,N+1$. 
%
Once the functional forms of $f$ and $h$ have been assigned, the MSF
associates to each value of the complex parameter $\alpha=\alpha^r +
j \alpha^i$, 
the maximum Lyapunov exponent of the system in each block. It can be
shown that for a large class of systems (in terms of the dynamic
function $f$ and the output function $h$), there exists a bounded
zone of the complex plane centered on the real axis, for which the
MSF is negative. 
Thus, the condition to be satisfied, in order to guarantee the
stability of the desired common solution $s(t)$, is that all the
$\sigma \mu_i$, $i=2,...,N+1$, belong to the bounded region of the
complex plane where the MSF is negative \cite{Pe:Ca}. Namely, for a
given form of the matrix $\mathcal{M}$, there is typically a finite
range of values of $\sigma$, say $\Sigma$, for which the stability
of the synchronous state can be achieved.

Hence, the spectrum of $\mathcal{M}$ affects the stability of the
synchronization manifold. In this paper, we focus on the case that
$\mathcal{M}$ has a real spectrum. Then, the lower the eigenratio
$R^{N+1}=\frac{\mu_{N+1}^r}{\mu_2^r}$, the larger the $\Sigma$
\cite{Pe:Ca}. More generally, when the Laplacian matrix 
has a complex spectrum, it has been proposed that the
width of $\Sigma$ can be evaluated in terms of both
$R^{N+1}=\frac{\mu_{N+1}^r}{\mu_2^r}$ and $M^{N+1}= \max_j \mu_j^i $
\cite{Bocc2}. Namely, the condition on the maximum imaginary part of
the spectrum gives information about the spread of the eigenvalues
along the direction of the imaginary axis; in the limit of
$M^{N+1}\rightarrow 0$, the entire spectrum tends to the real axis
and the (best) condition is recovered that $\Sigma$ depends only on
$R^{N+1}$.
Furthermore, as explained in \cite{Ni:Mo06}, the same approach 
can be extended also to the case of $\mathcal{M}$ being a
non-diagonalizable matrix, when the condition is satisfied that the
network embeds at least an oriented spanning tree \footnote{The
existence of an oriented spanning tree, having its root at the node
$N+1$, is guaranteed by matrix $\mathcal{A}$ to be irreducible.}.



%

Now, in the case often considered in the literature of all control
gains being the same, i.e. $\kappa=\kappa_1=\kappa_2=...=\kappa_n$,
by rewriting the system in (\ref{eq:net}) as
$\frac{dy_i}{dt}=f(y_i)-\kappa \sum_{j=1}^{N+1} {\mathcal M}'_{ij}
h(y_j)$, for $ i=1,...,N+1$, where
$\mathcal{M'}=\frac{\sigma}{\kappa}\mathcal{M}$, and by noticing
that the two matrices $\mathcal{M}$ and $\mathcal{M}'$ are
characterized by the same eigenratios, we obtain, for every choice
of $\sigma$, a similar condition on the interval of values of
$\kappa$, say $\mathcal{K}$, for which the reference evolution is
stable (this is true as long as the maximum imaginary part
$M^{N+1}$ remains negligible).

Note that, by taking this approach, we have succeeded in decoupling
the dynamical properties of the open-loop network (in terms of $f$
and $h$) from the factors encoded in the matrix $\mathcal{M}$.
Specifically: (i) the structural properties of the network, in terms
of the original network topology and the choice of the set of the
controlled nodes $\mathcal{C}$, and (ii) the choice for each node in
$\mathcal{C}$, $c_1, c_2, ..., c_n$  of the associated control gain
$\kappa_1, \kappa_2,...,\kappa_n$.

Thus, in analogy to the concept of network synchronizability defined in the literature \cite{Pe:Ca,Ni:Mo,report}, 
we propose to give a definition of the network
\emph{pinning-controllability} in terms of the widths of the ranges
$\Sigma$ and $\mathcal{K}$, for which the reference evolution
$y_1(t)=y_2(t)=...=y_N(t)=s(t)$ is stable. Specifically, the lower
the $R^{N+1}$ and $M^{N+1}$ are, the more the network is
pinning-controllable (note that this definition is independent of
the choice of the functions $f$ and $h$). Also, according to this
definition, it becomes possible to act on both the network topology
as well as the choice of the nodes to pin and their control gains,
in order to vary (and eventually improve) the network
pinning-controllability.

\section{Numerical Results}

\begin{figure}[!t]
  {\centerline{\psfig{figure=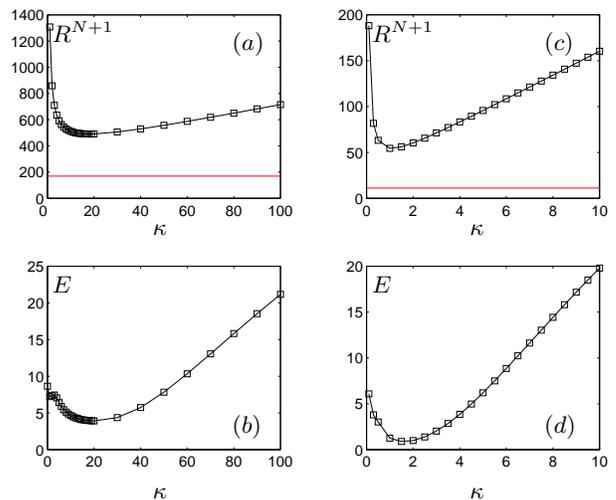,angle=0,width=8cm}}}
\begin{picture}(0,0)(0,0)
\put(-30,25){\small ${(b)}$} \put(88,25){\small ${(d)}$}
\put(-30,170){\small ${(a)}$} \put(88,170){\small ${(c)}$}
\put(-60,0){\small ${\kappa}$} \put(60,0){\small ${\kappa}$}
\put(-60,100){\small ${\kappa}$} \put(60,100){\small ${\kappa}$}
\put(-98,78){\small ${E}$} \put(23,78){\small ${E}$}
\put(-98,173){\small ${R^{N+1}}$} \put(23,173){\small ${R^{N+1}}$}
\end{picture}
\caption{\label{MAIN} \small \emph{Left pictures: Combinatorial
Laplacian (symmetric topology)}. (a) Behavior of the eigenratio
$\mu_{N+1}/\mu_2$ of the Laplacian spectrum as function of the
control gain $\kappa$ for a BA network, with average degree $\langle
k \rangle =2$, $p=0.1$, $\sigma=0.30$. (b) Control error at regime
$E$ as function of the control gain $\kappa$ under the same
conditions as in the upper plot.\\ \emph{Right pictures: Normalized
Laplacian (asymmetric topology)}. (c) $R^{N+1}$ vs. $\kappa$ for a
BA network, with average degree $\langle k \rangle =2$, $p=0.1$,
$\sigma=2.8$. (d) $E$ vs. $\kappa$ under the same conditions as in
the upper plot. In (a) and (c) the horizontal continuous lines
represent the eigenratio $R^{N}$ of the corresponding uncontrolled
networks.}
\end{figure}

\begin{figure*}[!t]
 {\centerline {\psfig{figure=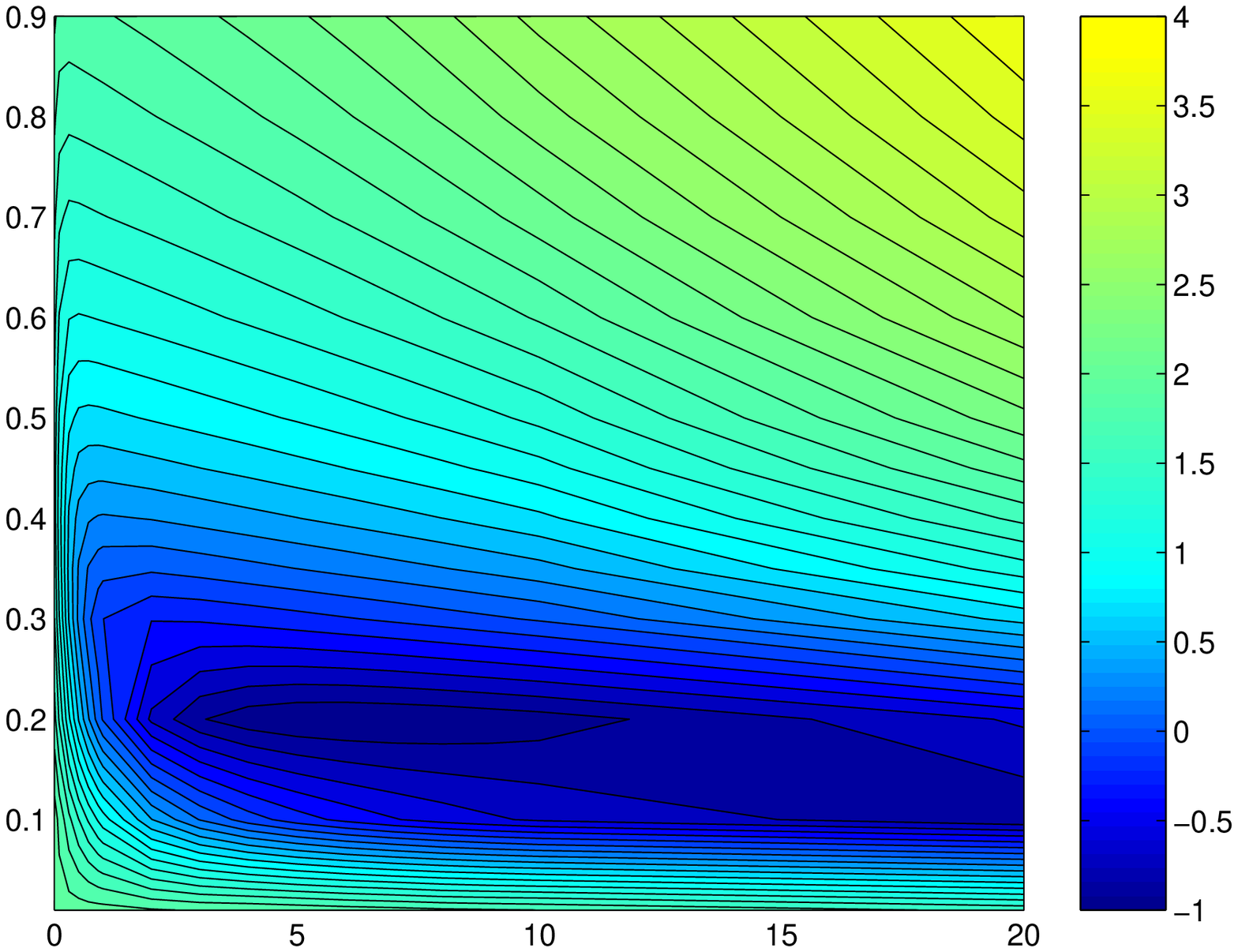,width=8cm}
 \quad \quad \quad \quad
 \psfig{figure=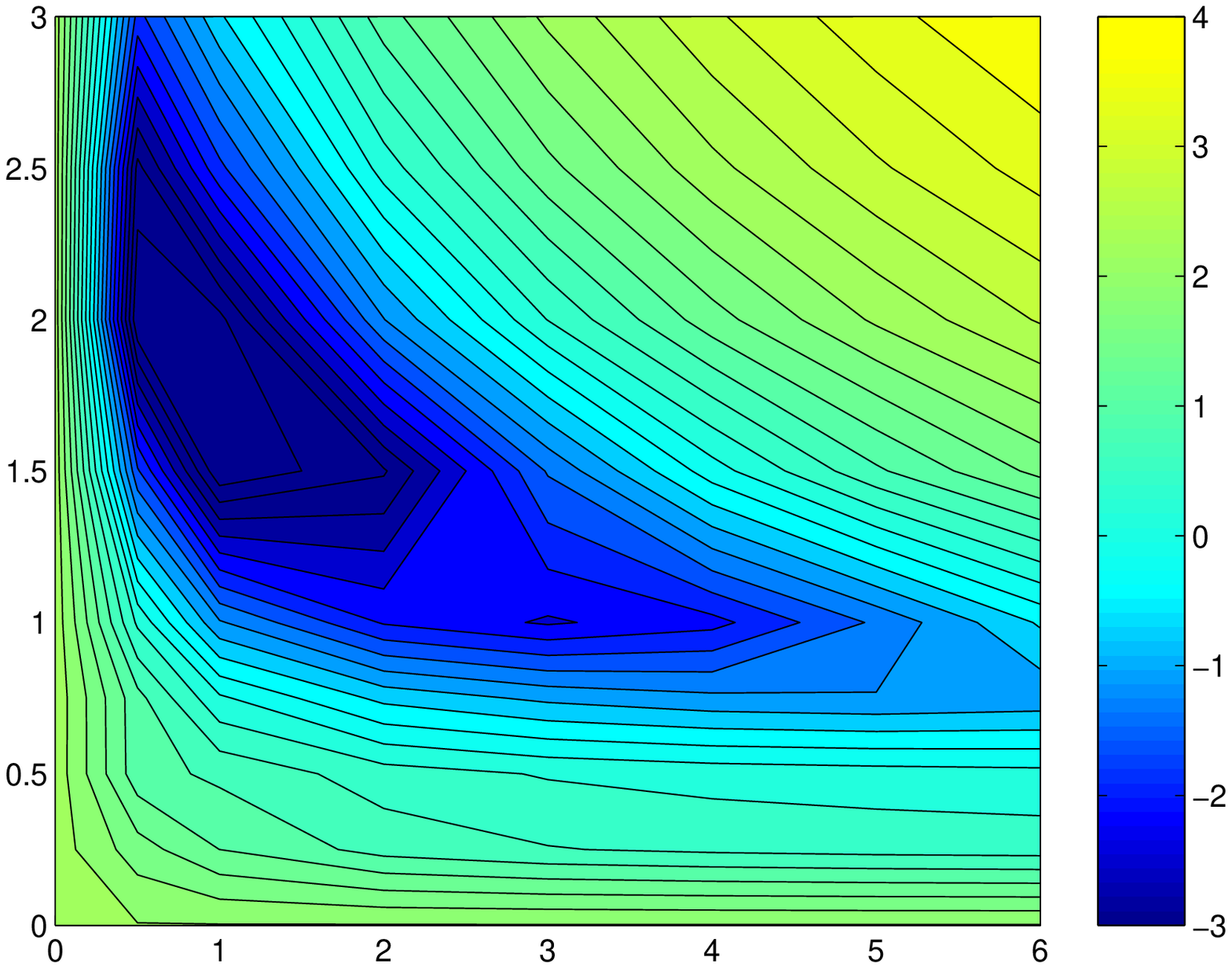,width=8cm}}}
\begin{picture}(0,0)(0,0)
\put(-155,0){\small ${\kappa}$} \put(-255,110){\small ${\sigma}$}
\put(125,0){\small ${\kappa}$} \put(13,110){\small ${\sigma}$}
\end{picture}
\caption{\label{CONTOURS} \small (Color online) Curve level sets of
the natural logarithm of $E$, asymptotic value of the control error,
as a function of both $\kappa$ and $\sigma$, for a BA network of
$10^3$ nodes, with average degree $\langle k \rangle =2$, $p=0.1$.
The left (right) panel shows the case of Combinatorial (Normalized)
Laplacian.} 
\end{figure*}

In this section, by following the approach presented above,
we present numerical evidence of the usefulness of $R^{N+1}$ 
as an index for evaluating the pinning-controllability of a given
complex network. Moreover, we show the behavior of $R^{N+1}$ under
variations of the control gain $\kappa$ and the pinning probability
$p$, in the case of scale-free complex networks.

To validate our theoretical findings, we consider the case of a
Barabasi-Albert (BA) scale-free network \cite{Ba:Al99},
characterized by a power-law degree distribution $P(k) \sim k^{-3}$
(we have checked our results to be reproduced in a similar qualitative way for different kinds of 
networks and lattices, including real-world samples of networks),
with $N$ identical R\"ossler oscillators placed at the network
vertices. Namely, the dynamics at each node $i$ is described by the
following vector field: $f(x_i)=f(x_{i1},x_{i2},x_{i3})=(- x_{i2}\;
-\; x_{i3}, \quad x_{i1}\;+\;0.165 x_{i2}, \quad
0.2+(x_{i1}-10)x_{i3})$. The output function $h$ has been chosen, as
in \cite{IJBCSI}, to be $h(x)=Hx$, where $H$ is the matrix: {\small
$\pmatrix{ (1 & 0 & 0) , (0 & 0 & 0),  (0 & 0 & 1) }$}, indicating
that the oscillators are coupled through the variables $x_{i1}$ and
$x_{i3}$  ($i=1,2,...,N+1$). 

Two cases have been considered: (1) the \emph{Combinatorial
Laplacian} defined as $\mathcal{L}_{ij}=\mathcal{L}_{ji}=-1$ if $i$
and $j$ are connected ($i \neq j$), $0$ otherwise, and
$\mathcal{L}_{ii}=\kappa_i \quad \forall i$ (corresponding to a
symmetric network); and (2) the \emph{Normalized Laplacian} defined
as $\mathcal{L}_{ij}=-1/\kappa_i$ if $i$ is connected to $j$ ($i
\neq j$), $0$ otherwise, and $\mathcal{L}_{ii}=1 \quad \forall i$,
which corresponds to an asymmetric network configuration (for more
details, see \cite{ChungBOOK}).

For simplicity, we evaluate here the effect of $n$ pinned nodes with
gains $\kappa_1=\kappa_2=...=\kappa_n=\kappa$, on the structural
parameters $R^{N+1}$ and $M^{N+1}$. For both the Combinatorial and
the Normalized Laplacians considered here, the spectra are real and
thus $M^{N+1}=0$ \footnote{Namely, in the case of Normalized
Laplacian, this can be shown by checking that there exists a
symmetric matrix similar to $\mathcal{M}$ \cite{Mo:Zh:Ku04}.}. The
main results are shown in Fig. \ref{MAIN}, for symmetric and
asymmetric topologies. In both cases, the eigenratio $R^{N+1}$ (in
Figs.\ref{MAIN}(a) and \ref{MAIN}(c)) is characterized by a minimum
around a specific value of the control
gain $\kappa$. 
This has immediate relevance as it suggests that by appropriately
tuning the control gain, it is possible to enhance the
pinning-controllability of the network. Specifically, we observe
that either a too large or too low value of $\kappa$ can reduce the
network pinning-controllability. 

As shown in Figs.\ref{MAIN}(a) and \ref{MAIN}(c), when $p$ is low ($p=0.1$ in the figure), $R^{N+1}$ is 
higher than the eigenratio $R^N$ associated with the corresponding
uncontrolled network (represented by the continuous horizontal lines
in Fig.\ref{MAIN}). Intuitively this is not surprising since, in
order to control the network, we are requiring \emph{one more}
eigenvalue (than in the case of generic network synchronization
\cite{Pe:Ca}) to fall
into the stability region of the MSF. 

Note that, for every choice of the coupling gain $\sigma$, the
eigenratio $R^{N+1}$ may be conveniently varied by choosing an
appropriate control gain $\kappa$. At the same time, for every
$\kappa$, a bounded region of stability can be defined on the
complex plane, and a timely choice of $\sigma$ is necessary in order
to place all the $\sigma \mu_i$ inside it. Thus, the stability of
the reference evolution is sensible to both the values of the
control gain and the coupling gain and values of $\kappa$ and
$\sigma$, which are either too large or too small, may prevent the
stability over the reference evolution of the network trajectories.

It is worth noting here that there is a prominent difference between
the effects that varying $\sigma$ and $\kappa$ can have on the
pinning-controllability of a given network. Specifically the width
of $\Sigma$ is determined essentially by the particular shape of the
MSF (which can be numerically computed, once the forms of the
functions $f$ and $h$ are given). On the other hand, the width of
$\mathcal{K}$ depends on the spectral properties of the matrix
associated to the extended network, defined as proposed in this
paper. Specifically, when the MSF is non-monotone, as in the case of
the
$xz$-coupled R\"ossler discussed in this paper, 
the width of the interval of the values of $\kappa$, for which the
stability of the reference evolution can be guaranteed, depends
simply on the eigenratio $R^{N+1}$, as shown in Fig.
\ref{MAIN}(a)-(c).

The asymptotic value of the control error $E=\frac{1}{(\Delta T) N}
\sum_{i=1}^{N} \int_T^{T+\Delta T} \| x_i(t)-s(t)\| dt$, with $\| x
\|= |x_{1}|+|x_{2}|+|x_{3}|$, has been computed under variations of
the control gain $\kappa$ in Figs. \ref{MAIN}(b) and \ref{MAIN}(d)
(respectively for fixed $\sigma=0.30$ and $2.8$)  and both $\kappa$
and $\sigma$ in Fig. \ref{CONTOURS}. Note that interestingly the
control error $E$ in Fig. \ref{CONTOURS} behaves locally as a
convex function of [$\kappa, \sigma$]. 

\begin{figure}[!t]
  {\centerline {\psfig{figure=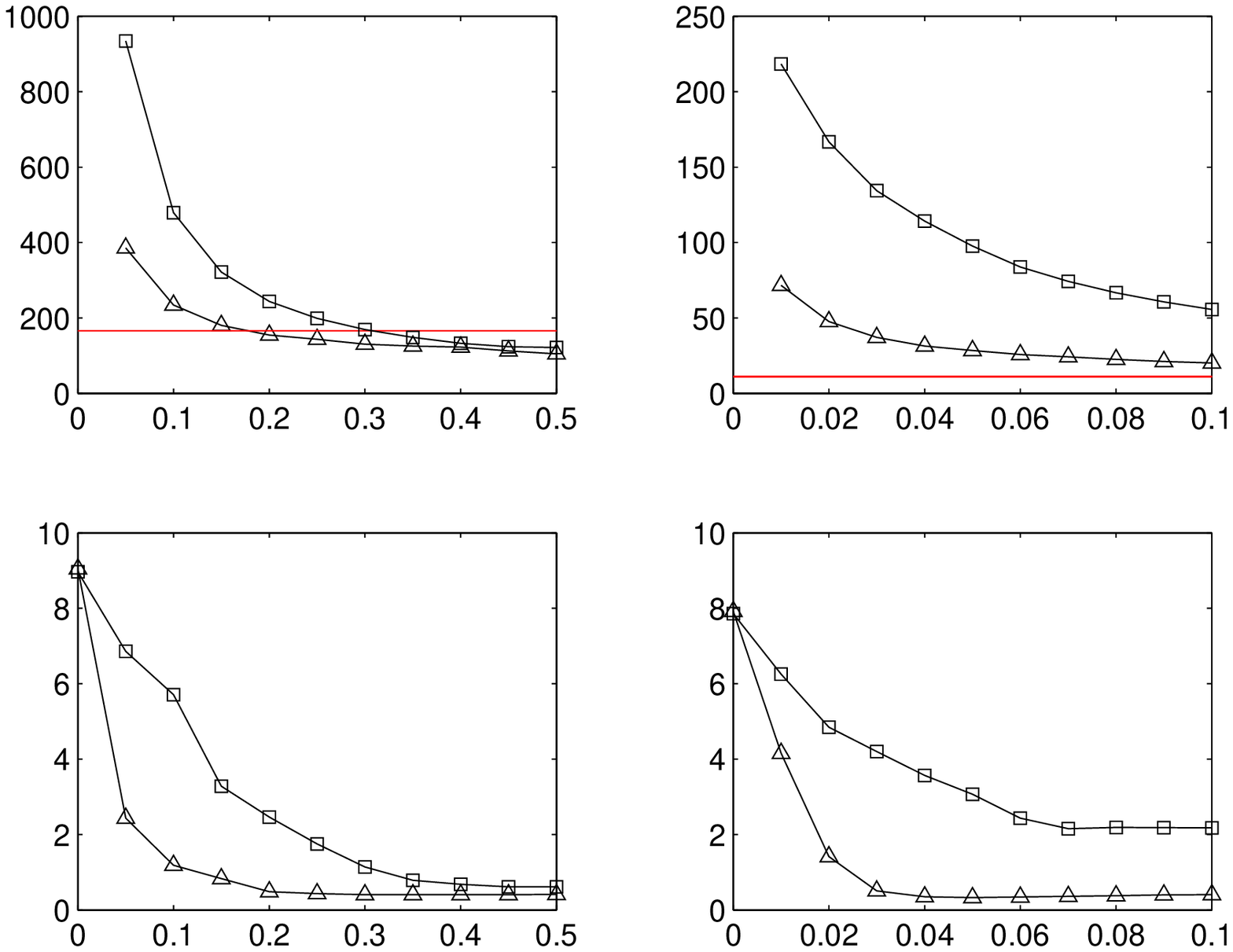,angle=0,width=8cm}}}
\begin{picture}(0,0)(0,0)
\put(-30,75){\small ${(b)}$} \put(95,75){\small ${(d)}$}
\put(-30,170){\small ${(a)}$} \put(95,170){\small ${(c)}$}
\put(-60,0){\small ${p}$} \put(60,0){\small ${p}$}
\put(-60,100){\small ${p}$} \put(60,100){\small ${p}$}
\put(-125,61){\small ${E}$} \put(0,61){\small ${E}$}
\put(-125,169){\small ${R^{N+1}}$} \put(-5,169){\small ${R^{N+1}}$}
\end{picture}
\caption{\label{P} \small  \emph{Left pictures: Combinatorial
Laplacian (symmetric topology)}. (a) Behavior of the eigenratio
$\mu_{N+1}/\mu_2$ of the Laplacian spectrum as function of the
control gain $\kappa$ for a BA network, with average degree $\langle
k \rangle =2$, $\sigma=0.30$, $\kappa=10$.  (b) Control error at
regime $E$ as function of the control
gain $\kappa$ under the same conditions as in the upper plot.\\
\emph{Right pictures: Normalized Laplacian (asymmetric topology)}.
(c) $R^{N+1}$ vs. $\kappa$ for a BA network, with average degree
$\langle k \rangle =2$, $\sigma=2.8$, $\kappa=1.5$. (d) $E$ vs.
$\kappa$ under the same conditions as in the upper plot. \\
In all the plots the legend is as
follows: squares (triangles) are used in the case 
of random (selective) pinning. In (a) and (c) the horizontal
continuous lines represent the eigenratio $R^{N}$ of the
corresponding uncontrolled networks.}
\end{figure}

Finally, the effect of a variable number of pinned nodes (randomly
selected within the network) is shown in Fig. \ref{P}.
Observe that, by increasing the number of reference nodes,
differently from the case of variable control gains, we recover a
monotone behavior in the eigenratio $R^{N+1}$, as shown in
Fig.\ref{P}(a). Interestingly, by increasing $p$, it is even
possible to make $R^{N+1}$ lower than $R^{N}$, i.e. controlling the
network becomes easier than synchronizing it. Also in Fig.
\ref{P}(b), the asymptotic value of the control error
$E$ is observed to decrease for increasing values of $p$. 

In Fig. \ref{P} a comparison between random pinning (squares) and
selective pinning (triangles) is also shown. In the case of
selective pinning, the nodes have been chosen in the order of
decreasing degree. We observe that selective pinning yields  a lower
eigenratio $R^{N+1}$ as well as a lower control error $E$ over
different values of $p$, as shown separately in Figs.
\ref{P}(a)-(b). This is in agreement with the previously reported
results where a different approach was used to evaluate the
effectiveness of pinning control schemes \cite{PinnA}. Moreover,
this shows the benefits of taking into account some network
topological features in choosing some suitable nodes to pin.

From a control design viewpoint, we observe that, while increasing
the control gains may lead to a loss of pinning-controllability of
the network, applying a larger number of controllers is always an
effective strategy. On the other hand, the requirement of pinning a
large number of nodes can make the technique unfeasible in those
networks where altering the dynamics of too many nodes can lead to a
loss of functionality of the network itself, or is too costly and
unpractical.

The case of Normalized Laplacian is shown in Figs. \ref{P}(c)-(d),
where the selective pinning is confirmed to perform better than the
random one. This indicates that high degree vertices continue to be
better suited to control the network than the others, even when the
total strength of the interaction at each vertex is rescaled  by its
degree.

Following these preliminary results, we believe other network
properties such as degree correlation \cite{Sorr:06}, clustering
\cite{Wa:St98}, centrality \cite{New:Bet} and community structure
\cite{Ne:Gi02} can influence the effectiveness of pinning-control
schemes. This represents the subject of future research activities.

\section{Conclusions}

In this paper we have presented a novel theoretical approach to
describing the controllability of networks under pinning control
schemes. We have defined the novel concept of
pinning-controllability and characterized the
pinning-controllability of a given network in terms of the coupling
gain, the control gain and the number of pinned nodes. We found that
this property can be analyzed by investigating the synchronizability
of an appropriately extended network. For instance, Fig. 1 shows
that there are values of the control gain which render the
eigenratio or the control error minimal. We wish to emphasize that
information such as this could be used to understand why some
physical networks possess certain values of the coupling and control
parameters. Moreover, the methodology presented here can be an
effective tool for the design of pinning control schemes of many
biological and technological networks.

Acknowledgments: This research was partially supported by the Hong
Kong Research Grants Council under the CERG Grant CityU1114/05E.


\end{document}